\documentclass{article}
\usepackage{amssymb}
\usepackage{amsmath}
\usepackage[dvips]{graphicx}
\usepackage{fleqn,ams}
\usepackage{verbatim}
\usepackage[cp1251]{inputenc}
\usepackage[dvips]{graphicx}
\begin{document}
\begin{center}
{\bf\Large Dynamic model of spherical perturbations in the Friedmann universe.
II. retarding solutions for the ultrarelativistic equation of state} \\[12pt]
Yu.G. Ignatyev, N. Elmakhi\\
Tatar State Humanitarity Pedagocical University\\ 1 Mezhlauk St.,
Kazan 420021, Russia
\end{center}

\begin{abstract}
Exact linear retarding spherically symmetric solutions of Einstein
equations linearized around Friedmann background for the
ultrarelativistic equation of state are obtained and investigated.
Uniqueness of the solutions in the $C^{1} $ class is proved.
\end{abstract}

\section{General  solution of evolutionary equation in the form
of power series and private cases.}

\subsection{Equations of spherical perturbations model}
As it is shown in Ref.[1], spherically symmetric perturbations of
Friedmann metrics are described by one scalar function $\delta \nu
(r,\eta $) connected with  perturbation of the metric tensor
component $g_{44} $ in the isotropic coordinates $(r,\eta )$ with
the relation
\begin{equation} \label{GrindEQ__1_}
\delta g_{44} =a^{2} (\eta )\delta \nu ,
\end{equation}
at that the representation turns out to be convenient

\begin{equation} \label{GrindEQ__2_}
\delta \nu =2\frac{\Phi (r,\eta )}{ar} =2\frac{\Psi (r,\eta )-\mu (\eta )}{ar} ,
\end{equation}
it allows to factor the solution into particlelike mode

\begin{equation} \label{GrindEQ__3_}
\delta \nu _{p} =-\frac{2\mu (\eta )}{ar} ,
\end{equation}
equivalent to Newton potential of the point mass $\mu (t)$, and the nonsingular mode

\begin{equation} \label{GrindEQ__4_}
\delta \nu _{0} =\frac{2\Psi (r,\eta )}{ar} ,
\end{equation}
the ``potential'' of which $\Psi (r,\eta )$ and its first
derivative by a radial variable at the beginning of the
coordinates satisfy the relations

\begin{equation} \label{GrindEQ__5_}
\mathop{\lim }\limits_{r\to 0} \left|\Psi (r,\eta
)\right|=]0,\quad \quad \mathop{\lim }\limits_{r\to 0}
\left|r\frac{\partial \Psi (r,\eta )}{\partial r} \right|=0.
\end{equation}
By a constant coefficient of barotrope $\kappa $ the mass of a
particle source $\mu (\eta )$ and the nonsingular mode potential
$\Psi (r,\eta )$ satisfy the evolutionary equations (75) and (84)
[1]:

\begin{equation} \label{GrindEQ__6_}
\ddot{\mu }+\frac{2}{\eta } \dot{\mu }-\frac{6(1+\kappa )}{(1+3\kappa )^{2} } \frac{\mu }{\eta ^{2} } =0,
\end{equation}

\begin{equation} \label{GrindEQ__7_}
\ddot{\Psi }+\frac{2}{\eta } \dot{\Psi }-\frac{6(1+\kappa )}{(1+3\kappa )^{2} } \frac{\Psi }{\eta ^{2} } -\kappa \Psi ''=0.
\end{equation}
In the  previous paper  [1] an exact solution of  the evolutionary
equation \eqref{GrindEQ__6_} for the particlelike  source mass was
also obtained

\begin{equation} \label{GrindEQ__8_}
\mu =\mu _{+} \eta ^{{\tfrac{2}{1+3\kappa }} } +
\mu _{-} \eta ^{-{\tfrac{3(1+\kappa )}{1+3\kappa }} } ,\quad \quad (1+\kappa )\ne 0;
\end{equation}

\begin{equation} \label{GrindEQ__9_}
\mu =\mu _{+} \eta +\frac{\mu _{-} }{\eta } ,\quad \quad (1+\kappa )=0.
\end{equation}
In  [1] it was noted  that the peculiarity in the solution
\eqref{GrindEQ__8_} by $1+3\kappa =0$ is coordinate disappearing
by coming from the temporal coordinate $\eta $ to the physical
time $t$

\begin{equation} \label{GrindEQ__10_}
t=\smallint a(\eta )d\eta ;\Rightarrow \quad t=\bar{c}
\eta ^{3{\tfrac{(1+\kappa )}{1+3\kappa }} } ;\quad \eta =
\mathop{\bar{C}}\nolimits_{1} t^{{\tfrac{1+3\kappa }{3(1+\kappa )}} } .
\end{equation}

Let us recall, the nonsingular part of  the energy density
perturbation $\delta \varepsilon (r,\eta )$ and the radial
velocity of  medium $v(r,\eta )$ are determined with help of the
potential function  $\Psi (r,\eta )$ by the relations

\begin{equation} \label{GrindEQ__11_}
\frac{\delta \varepsilon }{\varepsilon _{0} } =
-\frac{1}{4\pi ra^{3} \varepsilon _{0} } \left(3\frac{\dot{a}}{a} \dot{\Phi }-\Psi ''\right),
\end{equation}

\begin{equation} \label{GrindEQ__12_}
(1+\kappa )v=-\frac{1}{4\pi ra^{3} \varepsilon _{0} } \frac{\partial }{\partial r} \frac{\dot{\Phi }}{r} ,
\end{equation}
where $\varepsilon _{0} (\eta )$ is a nonperturbed energy density of Friedmann Universe?

\begin{equation} \label{GrindEQ__13_}
\varepsilon _{0} \sim \eta ^{-{\tfrac{6(1+\kappa )}{1+3\kappa }} }
; \quad a\sim \eta ^{{\tfrac{2}{1+3\kappa }} } ; \quad \varepsilon
_{0} a^{3} \sim \eta ^{-{\tfrac{6\kappa }{1+3\kappa }} } .
\end{equation}

In [1] by the variables separation method a general solution of
the evolutionary equation \eqref{GrindEQ__7_} satisfying the
conditions \eqref{GrindEQ__5_} in form of  an integral from Bessel
functions was also obtained. In this paper we shall obtain more
convenient solutions in form of power series.

\subsection{Class $C^{\infty } $ general solution in the
perturbation area; $\kappa \ne 0$, $1+\kappa \ne 0$}
As it was shown in the previous paper [1], the singular part of
the potential function $\delta \nu (r,\eta )$ is uniquely
extracted by the representation \eqref{GrindEQ__2_} in which the
potential function $\Psi (r,\eta )$ is nonsingular at the
beginning of the coordinates, that is satisfies the relations
\eqref{GrindEQ__5_} in  consequence of which, in particular

\begin{equation} \label{GrindEQ__14_}
\Psi (0,\eta )=0.
\end{equation}
Further supposing in the final neighborhood $r=0;r\in [0,r_{0})$
in which the metrics perturbation is localized, the potential
function $\Psi (r,\eta )$ belongs to the $C^{\infty } $class, let
us represent the solution of the  evolutionary equation
\eqref{GrindEQ__6_} satisfying the conditions \eqref{GrindEQ__5_}
in the form of  power series of the radial variable  $r$

\begin{equation} \label{GrindEQ__15_}
\Psi (r,\eta )=\sum _{n=1}^{\infty } \, \Psi _{n} (\eta )r^{n} .
\end{equation}
Let us underline the expansion \eqref{GrindEQ__15_} does not
consist of a member with a zero power $r$in consequence of the
relation \eqref{GrindEQ__14_}. Substituting the function $\Psi
(r,\eta ))$ in the form of the Eq. \eqref{GrindEQ__15_} into the
evolutionary equation \eqref{GrindEQ__6_} and setting the
coefficients at each power  $r$in the obtained equation equal to
zero, we get a chain of  linking equations

\begin{eqnarray} \label{GrindEQ__16_}
\Psi _{2m} =0;{\rm \; }\mathop{\ddot{\Psi }}\nolimits_{2m+1} +
2\frac{\mathop{\dot{\Psi }}\nolimits_{2m+1} }{\eta } -
\frac{6(1+\kappa )}{(1+3\kappa )^{2} } \frac{\Psi _{2m+1} }{\eta
^{2} } \nonumber\\= \kappa (2m+3)(2m+2)\Psi _{2m+3} ; \quad
m=\overline{0,\infty },\; (\kappa \ne 0,-1).
\end{eqnarray}
Thus, by $\kappa \ne 0$ the general solution of the evolutionary
equation \eqref{GrindEQ__6_} for the potential function $\Psi
(r,\eta )$ is a series by odd powers of the radial  variable $r$,
that is  by  $\kappa \ne 0,-1$ the potential function  $\Psi
(r,\eta )$ of the $C^{\infty } $ class is an odd function of the
radial variable $r$

\begin{equation} \label{GrindEQ__17_}
\Psi (r,\eta )=\sum _{p=0}^{\infty } \, \Psi _{2p+1} (\eta )r^{2p+1} .
\end{equation}

For the private particle solutions responding the specific
physical conditions the series can be broken at any odd  $n=N\ge
3$. In this case supposing for the last member of the series

\begin{equation} \label{GrindEQ__18_}
\Psi _{n} (\eta )=0;\quad n>N=2p+1,\quad (p=1,2...),
\end{equation}
we obtain from the Eq. \eqref{GrindEQ__16_}  a closed equation

\begin{equation} \label{GrindEQ__19_}
\mathop{\ddot{\Psi }}\nolimits_{2p+1} +2\frac{\mathop{\dot{\Psi }}\nolimits_{2p+1} }{\eta } -
\frac{6(1+\kappa )}{(1+3\kappa )^{2} } \frac{\Psi _{2p+1} }{\eta ^{2} } =0.
\end{equation}
As far as this equation does not differ from the evolutionary
equation for the particlelike source mass at all, its general
solution will coincide with the solution \eqref{GrindEQ__8_}
accurate within reterming

\begin{equation} \label{GrindEQ__20_}
\Psi _{2p+1} =C_{+}^{p} \eta ^{{\tfrac{2}{1+3\kappa }} } +C_{-}^{p} \eta ^{-{\tfrac{3(1+\kappa )}{1+3\kappa }} } ,
\end{equation}
where $C_{+}^{p} $ and $C_{-}^{p} $ are some constants.

Substituting this solution into the next to last  equation of the
chain \eqref{GrindEQ__16_} we shall get the equation for
determining $\Psi _{2p-1} $:

\begin{eqnarray} \label{GrindEQ__21_}
\mathop{\ddot{\Psi }}\nolimits_{2p-1} +2\frac{\mathop{\dot{\Psi
}}\nolimits_{2p-1} }{\eta } -\frac{6(1+\kappa )}{(1+3\kappa )^{2}
} \frac{\Psi _{2p-1} }{\eta ^{2} }\nonumber\\ =\kappa
(2p+3)(2p+2)\left[C_{+}^{p} \eta ^{{\tfrac{2}{1+3\kappa }} }
+C_{-}^{p} \eta ^{-{\tfrac{3(1+\kappa )}{1+3\kappa }} } \right].
\end{eqnarray}
In consequence on this equation linearity its general solution is a sum of the general solution of the corresponding homogeneous equation $\Psi _{2p-1}^{0} $ and a private solution of the inhomogeneous one $\Psi _{2p-1}^{1} $. But the general solution of the homogeneous one coincides with the already mentioned above solution  \eqref{GrindEQ__8_}, and the private solution can be introduced in the form of a sum of two solutions corresponding to the two members of the right part \eqref{GrindEQ__19_}. Therefore, evidently the corresponding private solution has the form

\begin{equation} \label{GrindEQ__22_}
\Psi _{2p-1}^{1} =A_{p-1} \eta ^{{\tfrac{2}{1+3\kappa }} +2} +B_{p-1} \eta ^{-{\tfrac{3(1+\kappa )}{1+3\kappa }} +2} .
\end{equation}
Thus, we find

\begin{equation} \label{GrindEQ__23_}
A_{p-1} =\frac{\kappa (1+3\kappa )}{2(7+9\kappa )} (2p+1)C_{+}^{p} ;\quad B_{p-1} =-\frac{\kappa (1+3\kappa )}{6(1-\kappa )} (2p+1)C_{-}^{p} .
\end{equation}
Substituting the obtained solutions into the previous equations let us repeat the analogues calculations/ Thus we pointed out the algorithm of constructing a general solution of an evolutionary equation  \eqref{GrindEQ__7_} reduced to repeating differentiation operations. This general solution corresponding to the highest power  $N=(2p+1)$ of the radial variable consists of  $2N$ arbitrary constants appearing every time by solving  the corresponding homogeneous differential equations.

\subsection{Case N=3}
Let us demonstrate the task solution in the simplest but as it
turns out the most important case when  $N=3$ ($p=1$) and the
series \eqref{GrindEQ__15_} consists of only two members
corresponding to the values$p=0,1$. In this case from Eqs.
\eqref{GrindEQ__15_}-\eqref{GrindEQ__23_} we find

\begin{equation} \label{GrindEQ__24_}
\Psi _{3} =C_{+}^{1} \eta ^{{\tfrac{2}{1+3\kappa }} } +C_{-}^{1} \eta ^{-{\tfrac{3(1+\kappa )}{1+3\kappa }} } ;
\end{equation}

\begin{eqnarray} \label{GrindEQ__25_}
\Psi _{1} =C_{+}^{0} \eta ^{{\tfrac{2}{1+3\kappa }} } +C_{-}^{0}
\eta ^{-{\tfrac{3(1+\kappa )}{1+3\kappa }} }+\nonumber\\
\frac{3\kappa (1+3\kappa )}{2(7+9\kappa )} C_{+}^{1} \eta
^{2{\tfrac{(2+3\kappa )}{1+3\kappa }} } -\frac{\kappa (1+3\kappa
)}{2(1-\kappa )} C_{-}^{1} \eta ^{-{\tfrac{1-3\kappa }{1+3\kappa
}} } .
\end{eqnarray}
Let us study now the specific cases  $\kappa =0$ and $1+\kappa =0$ which are out of the general solution \eqref{GrindEQ__15_}.

\subsection{Nonrelativistic matter $\kappa =0$}

\noindent

\noindent In this case we get from \eqref{GrindEQ__8_} the law of mass evolution of  a particlelike source (see Ref. [4])

\begin{equation} \label{GrindEQ__26_}
\mu =C_{+} \eta ^{2} +C_{-} \eta ^{-3} .
\end{equation}
The equation \eqref{GrindEQ__7_} for the potential function $\Psi $ takes the form

\begin{equation} \label{GrindEQ__27_}
\ddot{\Psi }+2\frac{\dot{\Psi }}{\eta } -6\Psi =0,\quad \kappa =0,
\end{equation}
whence we find

\begin{equation} \label{GrindEQ__28_}
\Psi =\phi _{+} (r)\eta ^{2} +\phi _{-} (r)\eta ^{-3} ,
\end{equation}
where  $\phi _{\pm } (r)$ are arbitrary functions $r$. In this case the mass of a particlelike source evolves according to the law

\begin{equation} \label{GrindEQ__29_}
\mu =\mu _{+} \eta ^{2} +\mu _{-} \eta ^{-3} .
\end{equation}

\subsection{Inflationary case  $\kappa +1=0$}
In this case the equation for the field function  F   is elliptic
\begin{equation} \label{GrindEQ__30_}
\ddot{\Phi }+2\Phi \frac{\dot{\Phi }}{\eta } +\Phi ''=0,\quad (\kappa +1=0),
\end{equation}
and the radial velocity of perturbation is not determined by the equation \eqref{GrindEQ__12_} which in this case gives

\begin{equation} \label{GrindEQ__31_}
\frac{\partial }{\partial \eta } \left(\frac{\Phi }{r} \right)=0,\quad 1+\kappa =0.
\end{equation}
Integrating \eqref{GrindEQ__31_} we find:

\begin{equation} \label{GrindEQ__32_}
\Phi =\phi (r)+\xi (\eta )r,
\end{equation}
where $\phi (r)$ and  $\xi (\eta )$ are some arbitrary functions of their arguments.
Substituting this solution into the equation \eqref{GrindEQ__31_} and
dividing the variables we find the function $\Phi (r)$

\begin{equation} \label{GrindEQ__33_}
\Phi =C_{1} -\frac{C_{2} }{2} r^{3} +r\left(C_{2} +\frac{C_{3} }{\eta } +C_{1} \eta \right),
\end{equation}
where $C_{1} $, $C_{2} $ and $C_{3} $ are arbitrary constants.

\section{Retarding spherical perturbations in ultrarelativistic
universe}

\subsection{Boundary conditions for the retarding solutions}
By  $\kappa >0$ The spherically symmetric perturbations of
Friedmann metrics are described by the hyperbolic equation, by
$\kappa <0$ - by elliptic one. Peculiarities of the hyperbolic
equations \eqref{GrindEQ__7_} are convergent and nonconvergent
waves

\begin{equation} \label{GrindEQ__34_}
r\mp \sqrt{\kappa } \eta ={\rm Const},
\end{equation}
spreading at the sound velocity

\begin{equation} \label{GrindEQ__35_}
v_{s} =\sqrt{\kappa } .
\end{equation}
Let us study the task for spherically symmetric solutions of
linearized Einstein equations  against the background of Friedmann
metrics with zero boarding conditions for the potential function
$\Phi (r,\eta )$ at the sound horizon corresponding to the
causality principle

\begin{equation} \label{GrindEQ__36_}
\Sigma :\quad r=r_{0} +\sqrt{\kappa } (\eta -\eta _{0} ).
\end{equation}
In terms of the introduced functions  $\Phi (r,\eta )$ and $\Psi (r,\eta $ these conditions can be written in the form

\begin{equation} \label{GrindEQ__37_}
\mathop{\left. \Phi (r,\eta )\right|}\nolimits_{r=r_{0} +\sqrt{\kappa } (\eta -\eta _{0} )} =0;\Leftrightarrow \mathop{\left. \Psi (r,\eta )\right|}\nolimits_{r=r_{0} +\sqrt{\kappa } (\eta -\eta _{0} )} =\mu (\eta );
\end{equation}

\begin{equation} \label{GrindEQ__38_}
\mathop{\left. \Phi '(r,\eta )\right|}\nolimits_{r=r_{0} +
\sqrt{\kappa } (\eta -\eta _{0} )} =0;
\Leftrightarrow \mathop{\left.
\Psi '(r,\eta )\right|}\nolimits_{rr_{0} +\sqrt{\kappa } (\eta -\eta _{0} )} =0.
\end{equation}

In this case in consequence of the equations for perturbations out
of the boundary of the sound horizon the perturbations of energy
density, pressure and velocity must automatically turn into
zero:\footnote{ At the very boundary of the sound front the energy
density perturbations and the velocities may practically have
final breaks.}

\begin{equation} \label{GrindEQ__39_}
\mathop{\left.\delta \varepsilon (r,\eta
)\right|}\nolimits_{r>r_{0} + \sqrt{\kappa } (\eta -\eta _{0} )}
=0; \quad \mathop{\left. v(r,\eta )\right|}\nolimits_{r>r_{0}
+\sqrt{\kappa } (\eta -\eta_{0} )} =0.
\end{equation}

An important private case of the boundary conditions
\eqref{GrindEQ__37_} and  \eqref{GrindEQ__38_} are the conditions
at ``the zero sound horizon ``

\begin{equation} \label{GrindEQ__40_}
\Sigma _{0} :\quad r=\sqrt{\kappa } \eta .
\end{equation}
$ $In this case instead of  \eqref{GrindEQ__37_} and \eqref{GrindEQ__38_} we have

\begin{equation} \label{GrindEQ__41_}
\mathop{\left. \Psi (r,\eta )\right|}\nolimits_{r\sqrt{\kappa } \eta } =
\mu (\eta );\quad \quad \mathop{\left. \Psi '(r,\eta )\right|}\nolimits_{r\sqrt{\kappa } \eta } =0.
\end{equation}

\subsection{Solutions with zero boundary conditions at zero
sound}

Let us study some private solutions in the form of power series by
radial variable satisfying the boundary conditions
\eqref{GrindEQ__41_}. Such perturbations can be generated by the
metrics fluctuations at zero moment of time and concentrated at
this moment of time in zero volume. Coming to the most convenient
radial variable ${\rm \varrho }$

\begin{equation} \label{GrindEQ__42_}
r=\sqrt{\kappa } \rho ,\quad (=\frac{1}{\sqrt{3} } \rho ),
\end{equation}
in the case of $\kappa =1/3$ let us write down the field equations
\eqref{GrindEQ__6_} and \eqref{GrindEQ__7_}. The dot denotes
derivatieves by anew temporal variable

\begin{equation} \label{GrindEQ__43_}
\ddot{\mu }+\frac{2}{\eta } \dot{\mu }-\frac{2\mu }{\eta ^{2} } =0,
\end{equation}

\begin{equation} \label{GrindEQ__44_}
\ddot{\Psi }+\frac{2}{\eta } \dot{\Psi }-\frac{2\Psi }{\eta ^{2} } -\Psi ''=0.
\end{equation}
then solving

\begin{equation} \label{GrindEQ__45_}
\mu =\mu _{+} \eta +\mu _{-} \eta ^{-2} ,
\end{equation}
where $\mu _{-} $ and $\mu _{+} $ are arbitrary constants.

\subsection*{N=3 }

As far as the Taylor-series expansion of the function $\Psi
$should not consist of the zero power  $r$ by definition and
therefore all the even powers in this expansion automatically
vanish, so setting further

\begin{equation} \label{GrindEQ__46_}
\Psi (\rho ,\eta )=\Psi _{1} (\eta )r+\Psi _{3} (\eta )r^{3}
\end{equation}
and dividing the variables in the equation \eqref{GrindEQ__44_} we get an equation for the functions$\Psi _{i} (\eta )$:

\begin{equation} \label{GrindEQ__47_}
\mathop{\ddot{\Psi }}\nolimits_{1} +{\tfrac{2}{\eta }} \mathop{\dot{\Psi }}\nolimits_{1} -{\tfrac{2\Psi _{1} }{\eta ^{2} }} =6\Psi _{3} ;
\end{equation}

\begin{equation} \label{GrindEQ__48_}
\mathop{\ddot{\Psi }}\nolimits_{3} +{\tfrac{2}{\eta }} \mathop{\dot{\Psi }}\nolimits_{3} -{\tfrac{2\Psi _{3} }{\eta ^{2} }} =0.
\end{equation}

In compliance with \eqref{GrindEQ__45_} we find from
\eqref{GrindEQ__48_}

\begin{equation} \label{GrindEQ__49_}
\Psi _{3} =C_{3}^{+} \eta +C_{3}^{-} \eta ^{-2} ,
\end{equation}
where $C_{\pm }^{3} $ are some constants? Substituting the
solution \eqref{GrindEQ__49_} into the right part of the equation
\eqref{GrindEQ__47_} and defining private solutions of the
obtained equation we shall find its general solution

\begin{equation} \label{GrindEQ__50_}
\Psi _{1} =C_{1}^{+} \eta +C_{1}^{-} \eta ^{-2} +\frac{3}{5} C_{3}^{+} \eta ^{3} -3C_{3}^{-} .
\end{equation}
Thus,

\begin{equation} \label{GrindEQ__51_}
\Psi =(C_{1}^{+} \eta +C_{1}^{-} \eta ^{-2} +\frac{3}{5} C_{3}^{+} \eta ^{3} -3C_{3}^{-} )r+(C_{3}^{+} \eta +C_{3}^{-} \eta ^{-2} )r^{3} .
\end{equation}
Further calculating $\Psi (\rho ,\eta )_{|r=\eta } \equiv \Psi (\eta ,\eta )$ ,
substituting the result into the boundary condition   \eqref{GrindEQ__41_}
and setting equal the coefficients in the obtained equation by simultaneous
powers $\eta $ we shall get a set of equations for the constants $C_{i}^{\pm } ,\mu _{\pm } $

\begin{equation} \label{GrindEQ__52_}
\left. \begin{array}{l} {\eta ^{-2} } \\ {\eta ^{-1} } \\ {\eta } \\ {\eta ^{2} }
\\ {\eta ^{4} } \end{array}\right|\begin{array}{l} {\begin{array}{l} {0{\rm \; \; \;
\; \; \; \; }=\mu _{-} ;} \\ {C_{1}^{-} {\rm \; \; \; \; }=0;} \\ {-2C_{3}^{-} =\mu _{+} ;} \end{array}} \\ {\begin{array}{l} {C_{1}^{+} {\rm \; \; \; \; }=0;} \\ {\frac{8}{5} C_{3}^{+} {\rm \; }=0.} \end{array}} \end{array}
\end{equation}
Thus, there are only two nonzero constants  included into the
expression for  $\nu $: $\mu _{+} $ and $C_{3}^{-} =-1/2\mu _{+}
$. So, finally
\begin{equation} \label{GrindEQ__53_}
\Psi (\rho ,\eta )=\left\{\frac{3}{2} \mu _{+} \rho -\frac{1}{2} \mu _{+} \frac{\rho ^{3} }{\eta ^{2} } ,\_ \_ r\eta ;\_ \_ \mu _{+} \eta ,\_ \_ \rho >\eta \right. \Rightarrow
\end{equation}

\[\Phi (\rho ,\eta )=\left(\frac{3}{2} \mu _{+} \rho -\frac{1}{2}
\mu _{+} \frac{\rho ^{3} }{\eta ^{2} } \right)\chi (\eta -\rho ),-\]
It is the obtained previously retarding solution with zero
boundary conditions at zero sound horizon [2] and$\chi (z)$ is the
Heavyside function. At that supposing the following for the
ultrarelativistic equation of state

\begin{equation} \label{GrindEQ__54_}
a(\eta )=\eta ;\quad \quad (\kappa =1/3),
\end{equation}
we obtain

\begin{equation} \label{GrindEQ__55_}
\nu (\rho ,\eta )=\left(3\frac{\mu _{+} }{\eta } -
\mu _{+} \frac{\rho ^{2} }{\eta ^{3} } -2\frac{\mu _{+} }{\rho } \right)\chi (\eta -\rho ),
\end{equation}
It is easy to see that in this case at the surface of zero sound
front $\rho =\eta $ the boundary conditions \eqref{GrindEQ__41_}
are fulfilled identically.

\subsection*{N=5}

Now let us study the fifth degree multinomial as a solution. In
this case instead of the relations \eqref{GrindEQ__44_} and
\eqref{GrindEQ__45_} we have

\begin{equation} \label{GrindEQ__56_}
\Psi (\rho ,\eta )=\Psi _{1} (\eta )\rho +\Psi _{3} (\eta )\rho ^{3} +
\Psi _{5} (\eta )\rho ^{5} ;
\end{equation}
\begin{equation} \label{GrindEQ__57_}
\mathop{\ddot{\Psi }}\nolimits_{1} +{\tfrac{2}{\eta }}
\mathop{\dot{\Psi }}\nolimits_{1} -{\tfrac{2\Psi _{1} }{\eta ^{2} }} =6\Psi _{3} ;
\end{equation}

\begin{equation} \label{GrindEQ__58_}
\mathop{\ddot{\Psi }}\nolimits_{3} +{\tfrac{2}{\eta }}
\mathop{\dot{\Psi }}\nolimits_{3} -{\tfrac{2\Psi _{3} }{\eta ^{2} }} =20\Psi _{5} ;
\end{equation}

\begin{equation} \label{GrindEQ__59_}
\mathop{\ddot{\Psi }}\nolimits_{5} +{\tfrac{2}{\eta }}
\mathop{\dot{\Psi }}\nolimits_{5} -{\tfrac{2\Psi _{5} }{\eta ^{2} }} =0.\begin{array}{l} {} \\ {} \end{array}
\end{equation}
Similarly to the previous case we have

\begin{equation} \label{GrindEQ__60_}
\Psi _{5} =C_{5}^{+} \eta +C_{5}^{-} \eta ^{-2} .
\end{equation}

Substituting \eqref{GrindEQ__60_} into the right part of the
equation \eqref{GrindEQ__58_}, solving it similarly to the
previous case and substituting the obtained solution into the
equation \eqref{GrindEQ__57_} we get finally

\begin{equation} \label{GrindEQ__61_}
\Psi _{3} =C_{3}^{+} \eta +C_{3}^{-} \eta ^{-2} +2C_{5}^{+} \eta ^{3} -10C_{5}^{-} ;
\end{equation}

\begin{equation} \label{GrindEQ__62_}
\Psi _{1} =C_{1}^{+} \eta +C_{1}^{-} \eta ^{-2} +\frac{3}{5} C_{3}^{+} \eta ^{3} -
3C_{3}^{-} +\frac{3}{7} C_{5}^{+} \eta ^{5} -15C_{5}^{-} \eta ^{2} ;
\end{equation}
and thus

\[\Psi (\rho ,\eta )=(C_{1}^{+} \eta +C_{1}^{-} \eta ^{-2} +
\frac{3}{5} C_{3}^{+} \eta ^{3} -3C_{3}^{-} +\frac{3}{7} C_{5}^{+} \eta ^{5} -15C_{5}^{-} \eta ^{2} )r+\]

\begin{equation} \label{GrindEQ__63_}
+(C_{3}^{+} \eta +C_{3}^{-} \eta ^{-2} +2C_{5}^{+} \eta ^{3} -
10C_{5}^{-} )r^{3} +(C_{5}^{+} \eta +C_{5}^{-} \eta ^{-2} )r^{5} .
\end{equation}

Substituting the equation \eqref{GrindEQ__63_} into the boundary
conditions \eqref{GrindEQ__41_} and equating the coefficients
under equal  $\eta $ degrees we get equations for the constants
$C_{i}^{\pm } ,\mu _{\pm } $

\begin{equation} \label{GrindEQ__64_}
\begin{array}{l|lll} {\eta ^{-2} } & {0} & {=} & {\mu _{-} ;} \\
{\eta ^{-1} } & {C_{1}^{-} } & {=} & {0;} \\ {\eta } & {-2C_{3}^{-} } & {=}
& {\mu _{+} ;} \\ {\eta ^{2} } & {C_{1}^{+} } & {=} & {0;} \\ {\eta ^{3} }
& {-24C_{5}^{-} } & {=} & {0;} \\ {\eta ^{4} } & {\frac{8}{5} C_{3}^{+} }
& {=} & {0;} \\ {\eta ^{6} } & {\frac{24}{7} C_{5}^{+} } & {=} & {0.} \end{array}
\end{equation}
Thus,

\begin{equation} \label{GrindEQ__65_}
C_{5}^{\pm } =0,
\end{equation}
and the rest constants values coincide with the obtained ones
above. It means that the solution coincides with the obtained one
above. It is easy to show, adding of any new members of the series
does not change the situation.

Thus, we have proved the theorem: \vspace{12pt}

\textbf{Theorem 1.} {\sl The only spherically-symmetric solution
of the $C^{1} $ class of linearized around Friedmann space-plane
solution to the Einstein equations for an ideal ultrarelativistic
fluid, corresponding to the zero boundary conditions at the zero
sound horizon \eqref{GrindEQ__41_}, is the solution
\eqref{GrindEQ__53_} (it is equivalent to (55))}.

\subsection{Investigation of the retarding solution}

From the formula \eqref{GrindEQ__55_} one can immediately see
continuity of not only the first radial derivatives but also the
first temporal ones of the $\delta \nu$ metrics perturbation and
the potential functions

\begin{equation} \label{GrindEQ__66_}
\mathop{\left. \frac{\partial \delta \nu (\rho ,\eta )}{\partial \rho }
\right|}\nolimits_{\rho =\eta } =0;
\end{equation}

\begin{equation} \label{GrindEQ__67_}
\mathop{\left. \frac{\partial \delta \nu (\rho ,\eta )}{\partial \eta } \right|}\nolimits_{\rho =\eta } =0;
\end{equation}
In Fig.1 and 2 graphs of temporal evolution of the potential
functions $\Phi (\rho ,\eta )$ and  $-\delta \nu (\rho ,\eta )$
are shown

\noindent

\begin{flushleft}
\begin{tabular}{ll}
\includegraphics[width=57mm, height=57mm,angle=-90]{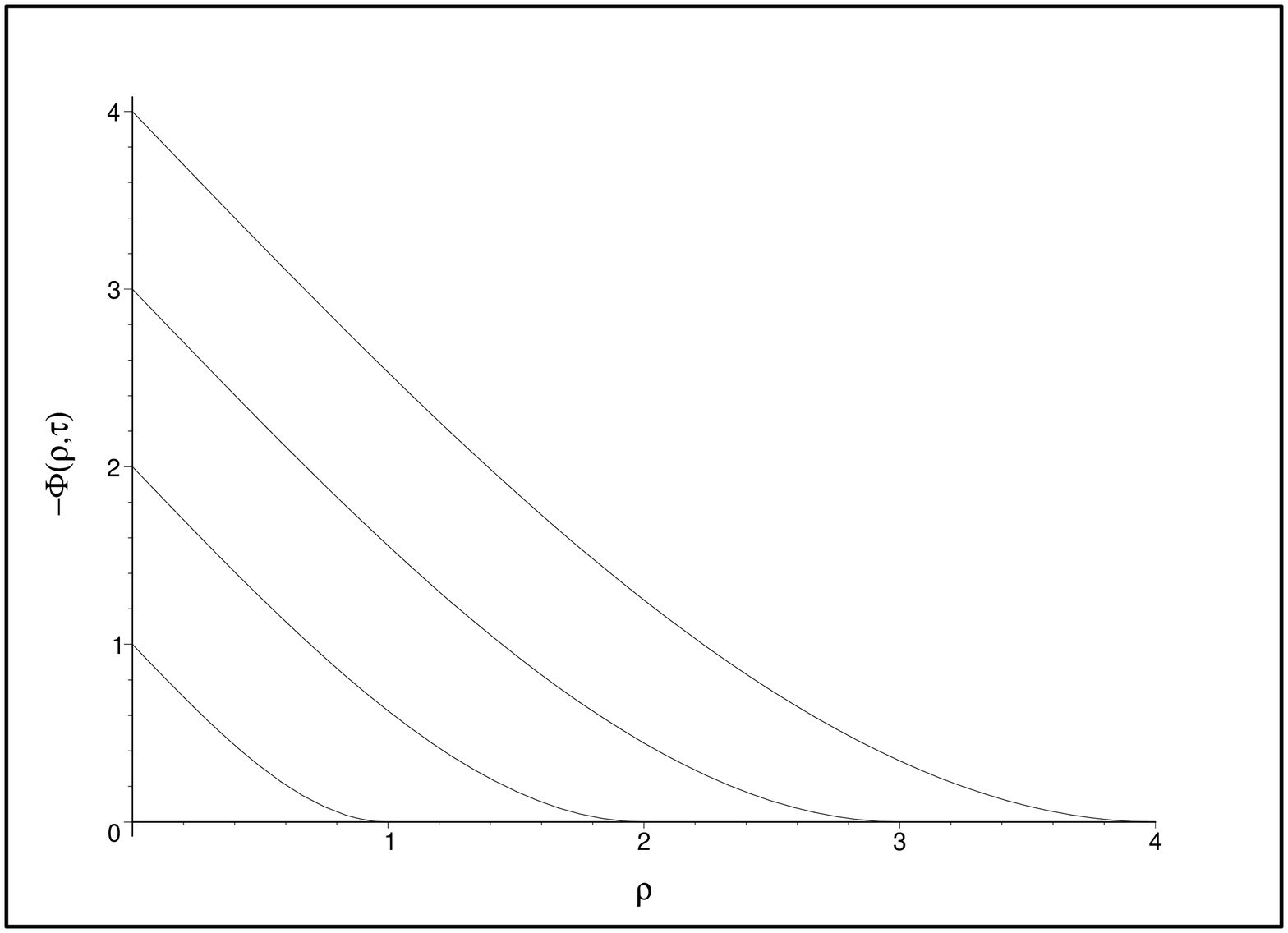}
& \includegraphics[width=57mm, height=57mm,angle=-90]{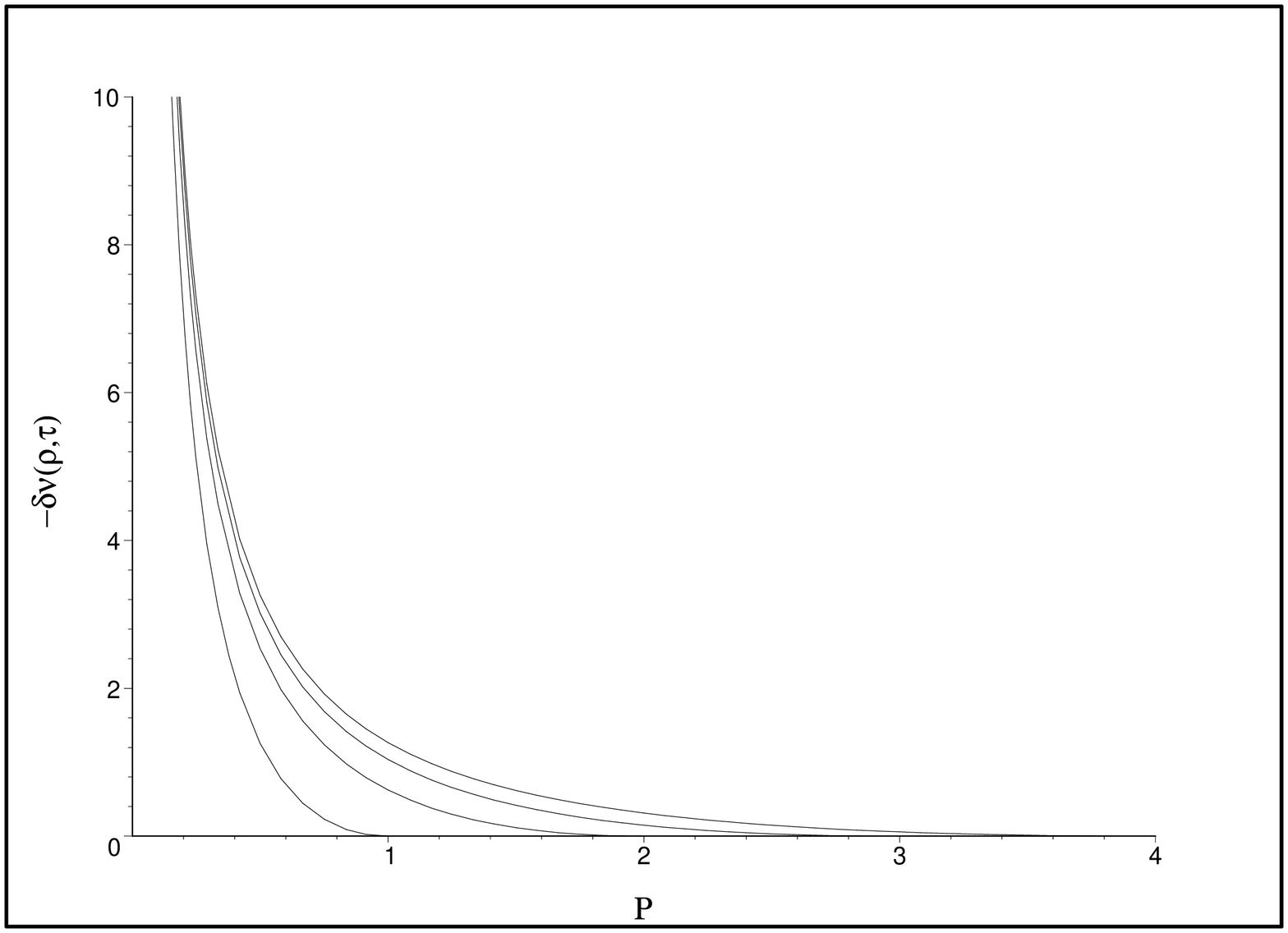} \\[24pt]
\parbox{57mm}{\vspace{12pt}\textbf{Fig.1}. Evolution of the potential function $-\Phi
(\rho ,\eta )$. Bottom-up $\eta =1;2;3;4$. Along the abscissa axis
the radial variable values $\rho =r\sqrt{3} $ are laid off.}   &
\parbox{57mm}{\vspace{12pt}\textbf{Fig 2.}  Evolution of the metrics perturbation
$-\delta \nu (\rho ,\eta )$. Bottom-up $\eta =1;2;3;4$. Along the
abscissa axis the radial variable values  $\rho =r\sqrt{3} $ are
laid off.}
\\
\end{tabular}
\end{flushleft}

\noindent
The second radial derivatives of the potential functions
and metrics have a final break at the sound horizon. In
consequence of this the energy density perturbation has a final
break at the sound horizon also. Calculating the relative energy
density of the spherical perturbations according to the formula
\eqref{GrindEQ__11_} we find
\begin{equation} \label{GrindEQ__68_}
\frac{\delta \varepsilon }{\varepsilon _{0} } =
-\frac{3\sqrt{3} \mu _{+} }{4\pi \rho } \left(\frac{\rho ^{3} }{\eta ^{3} } +3\frac{\rho }{\eta } -1\right).
\end{equation}
The jump of the relative energy density at the sound horizon is
\begin{equation} \label{GrindEQ__69_}
\Delta =\mathop{\left. \frac{\delta \varepsilon }{\varepsilon _{0} } \right|}\nolimits_{\rho =
\eta } =-\frac{9\sqrt{3} \mu _{+} }{4\pi \eta }
\end{equation}
and it decreases by the time (see Fig. 3). Let us turn to the
formula for the radial velocity \eqref{GrindEQ__12_}. Substituting
the equation for $\Phi $ into this formula and coming to the
radial variable $\rho $ we obtain

\begin{equation} \label{GrindEQ__70_}
v=-\frac{9\sqrt{3} \mu _{+} \eta }{16\pi \rho ^{3} } \left(1+2\frac{\rho ^{3} }{\eta ^{3} } \right).
\end{equation}
Evolution of the perturbation radial velocity is shown in Fig.4.

\begin{flushleft}
\begin{tabular}{ll}
\includegraphics[width=57mm, height=57mm,angle=-90]{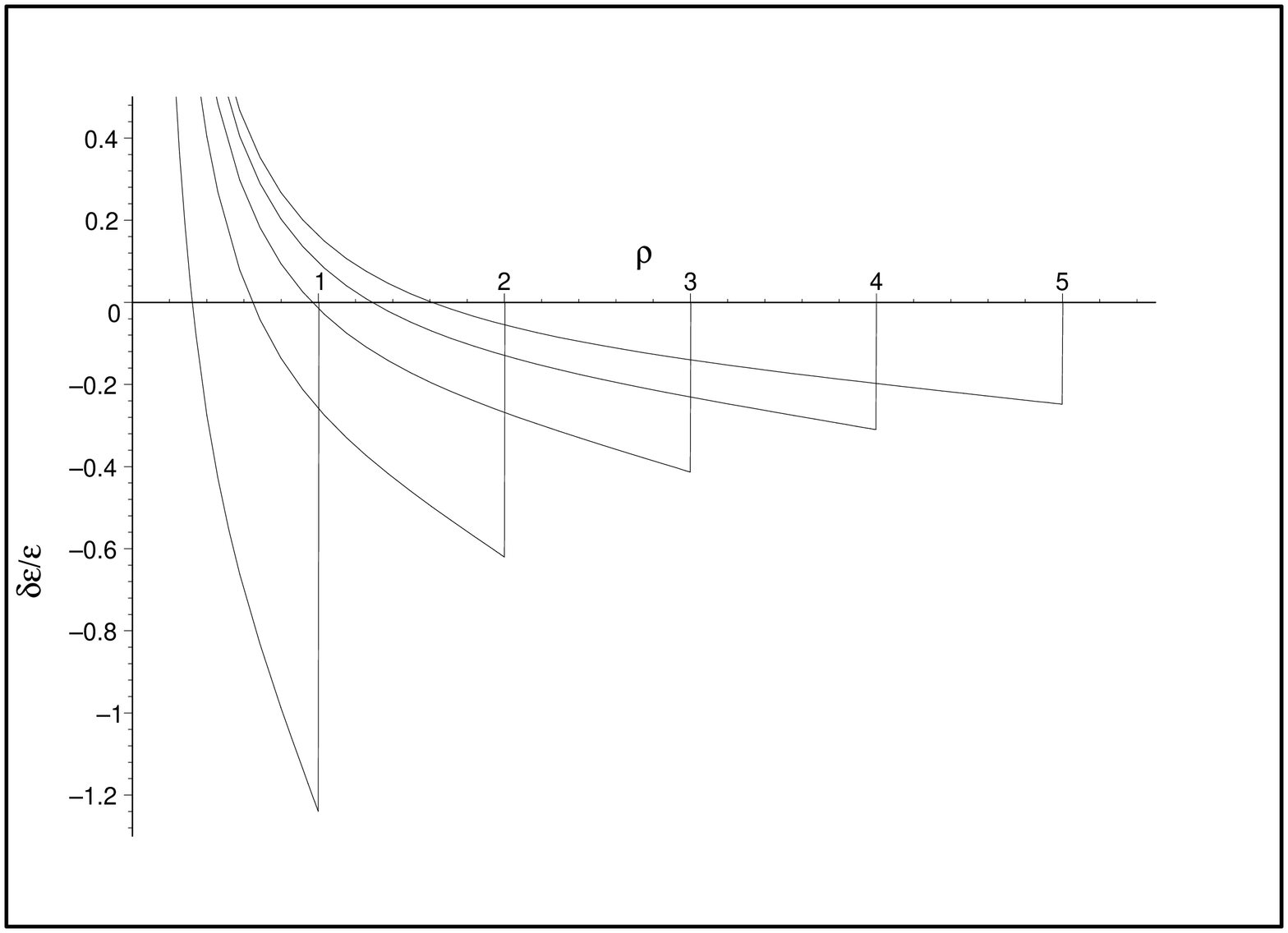}
& \includegraphics[width=57mm, height=57mm,angle=-90]{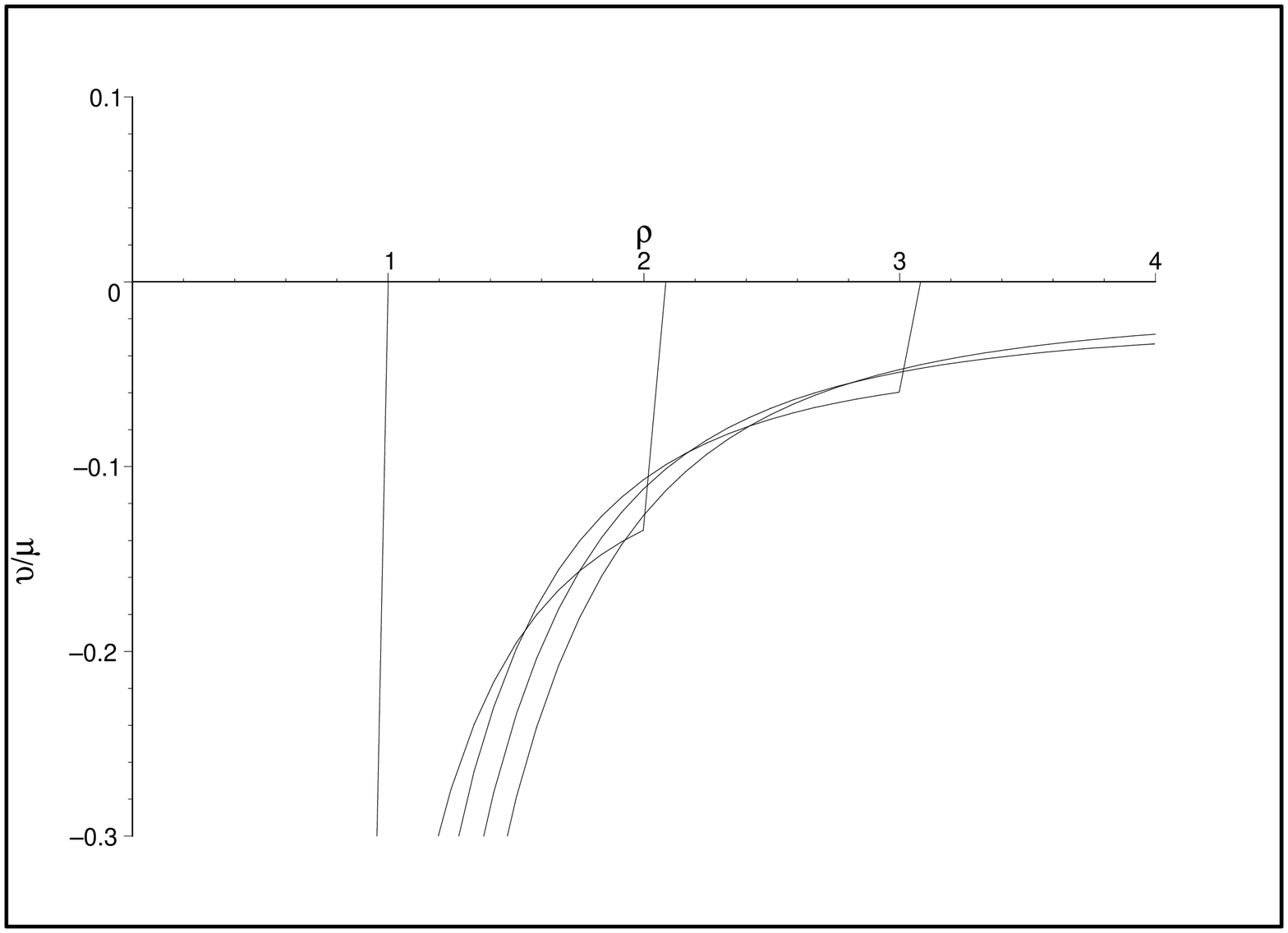} \\[12pt]
\parbox{57mm}{\vspace{12pt}\textbf{Fig.3}. Evolution of the relative perturbation of the energy density
$\delta \varepsilon /\varepsilon _{0} $. From  left to  right
$\eta =1;2;3;4;5$. Along the abscissa axis the radial variable
values $\rho =r\sqrt{3} $ are laid off.}  &
\parbox{57mm}{\vspace{12pt}\textbf{Fig 4.}  Evolution of the given radial velocity  $v(\rho ,\eta )/\mu _{+} $.
From left to right  $\eta =1;2;3;4;5$. Along the abscissa axis the
radial variable values  $\rho =r\sqrt{3} $are laid off.}
\\
\end{tabular}
\end{flushleft}

\section{Conclusion}

Summarizing let us point out the following. It is easy to see that
the obtained general retarding solution with a central singular
source coincides with the earlier obtained ones [2],[3],[4]and
[5]. However, note that in the mentioned works the solutions
\eqref{GrindEQ__53_}-\eqref{GrindEQ__55_} were pointed out as
private retarding solutions. The obtained retarding solutions
belong to the class $C^{1} $, however the second derivatives by
radial variable have a break of the first genus at the sound
horizon.  As a result the energy density break at the sound
horizon corresponds to the obtained retarding solution.  Because
of the pointed out conditions to find out the nature of the
density break it is necessary to solve two tasks: to find the
retarding solution for the barotrope arbitrary coefficient and
solve the Cauchy task for the initially localized perturbation. We
shall solve these tasks in our next papers.

\end{document}